\documentclass[twocolumn]{webofc}
\usepackage[varg]{txfonts}   % Web of Conferences font
\begin{document}
\title{Investigation of magnetocaloric effect: Stoner approximation vs DMFT}

\author{\firstname{Petr\ A.} \lastname{Igoshev}\inst{1}\fnsep\thanks{\email{igoshev_pa@imp.uran.ru}} \and
        \firstname{Igor\ A.} \lastname{Nekrasov}\inst{2}\fnsep\thanks{\email{nekrasov@iep.uran.ru}} \and
        \firstname{Nikita\ S.} \lastname{Pavlov}\inst{2}\fnsep\thanks{\email{pavlov@iep.uran.ru}} \and
        \firstname{Timur\ H.} \lastname{Chinyaev}\inst{2}       \and 
        \firstname{Edward\ O.} \lastname{Yakupov}\inst{3}
}

\institute{M.N. Mikheev Institute for Metal Physics, Russian Academy of Sciences, Ural Branch,
	S.Kovalevskoi str. 18, Ekaterinburg, 620290, Russia 
\and
        Institute for Electrophysics, Russian Academy of Sciences, 
        Ural Branch, Amundsen str. 106,  Ekaterinburg, 620016, Russia        
\and
        P. N. Lebedev Physical Institute, Russian Academy of Sciences, Moscow, 119991, Russia
          }

\abstract{
A comparative study of the magnetocaloric effect (MCE) in metals within the single-band Hubbard model on the face-centered cubic
(fcc) lattice using both mean-field (Stoner) approximation (MFA) and dynamical mean-field theory (DMFT) is done. The MCE is investigated in the case of second order magnetic phase transition from ferromagnet to paramagnet. To ensure
presence of itinerant ferromagnetism in the Hubbard model the special case of spectrum parameters generating giant van Hove singularity at the bottom of the band is considered, while the Fermi level $E_{\rm f}$ is in the vinicity of the band bottom.  To compare MCE within MFA and DMFT  temperature dependence of magnetization, total energy and finally entropy for a set of Coulomb interactions $U$ at zero and finite values of magnetic field $h$ for both methods were performed. Also one of the MCE
potentials, isothermal entropy change, as a function of temperature $\Delta S (T)$ for both MFA and DMFT is calculated. In the
MFA, the expected maximum value of $\Delta S (T)$ at the Curie temperature $T_C$ ($\Delta S_{\rm max}$) quite significantly decreases
while $U$ grows. Similar but much weaker decreasing of $\Delta S_{\rm max}$ is found for DMFT results.
The account of local quantum fluctuations results in larger values of $\Delta S_{\rm max}$ within DMFT than within MFA. A peak width of $\Delta S (T)$ at half height is approximately the same for both methods. 
Another effect of DMFT local quantum fluctuations is the destruction of 
anomalous Curie temperature $T_C$ dependence on $U$ present in MFA, which is invoked by an effect of giant van Hove singularity. 
However the relative cooling power (RCP) is very close in DMFT and MFA for the same model parameters and goes down upon $U$ increase.
}
\maketitle
%
%====================================================================================
\section{Introduction}
\label{sec:intro}

The investigation of the magnetocaloric effect (MCE) accumulates a huge amount of experimental data~(see reviews~\cite{Gschneidner2005,Franco2012,Li2016}). 
Theoretical approaches to the study of MCE are described in the review~\cite{Oliveira2010}.
Appropriate materials for using as a working body for magnetic cooling are discussed in Refs.~\cite{Tishin2007, Tishin2014}. Application of the MCE in medicine is reviewed in Ref.~\cite{Tishin2016}.

In our group, a series of articles on theoretical investigation of the MCE was carried out.
In particular, MCE was investigated for Heisenberg model within mean-field and random phase approximations \cite{Kokorina2013}.
MCE in presence of different types of anisotropy for antiferromagnets \cite{KO_2014_2,KO_2016} and paramagnets with non-Kramers ions \cite{Kokorina2017}
and Van Vleck paramagnet \cite{Kokorina2018_2} and Ising ferromagnet \cite{Kokorina2018_1} was investigated. Also MCE for some exotic systems was studied \cite{zarubin2015}.

Single-band Hubbard model in presence of magnetic field has a Hamiltonian
\begin{equation}
	H = \sum_{ij\sigma}(t_{ij} - h\delta_{ij})c^\dag_{i\sigma}c_{j\sigma} + U\sum_i c^\dag_{i\uparrow}c_{i\uparrow}c^\dag_{i\downarrow}c_{i\downarrow},
\end{equation}
where $c^\dag_{i\sigma}/c_{i\sigma}$ is electron creation/annihilation operator, $\sigma$ is spin projection, $U$ is local Coulomb interaction parameter, $t_{ij} = -t(t')$ for nearest (next-nearest) neighbor hopping transfer integral.
Recently investigation of of MCE for the Hubbard model within the mean-field (Stoner) approximation (MFA) for the second~\cite{Igoshev2017} and first~\cite{Igoshev2019} order magnetic phase transitions was carried out. Next natural step is to take into account the electron-electronic correlations more accurately within dynamical mean-field theory (DMFT)~\cite{Georges1996}.

In this paper we present MCE isothermal entropy change $\Delta S$ as a function of temperature and on-site Coulomb interaction $U$ obtained for single-band Hubbard model within DMFT in comparison with mean-field (Stoner) approximation. In section~\ref{sec:theory} we present the technique for calculation of the isothermal entropy change $\Delta S$ within the DMFT framework. In section~\ref{sec:technical} technical details are presented. Further, in section~\ref{sec:results} DMFT and MFA results of $\Delta S$ calculation are discussed.

%====================================================================================
\section{Entropy calculation technique}%Computational details}
\label{sec:theory}
To calculate isothermal entropy change $\Delta S$ within MCE we start with definition of total energy of $N$ site lattice interacting system:
\begin{equation}
d\mathcal{E}_{N} = TdS_{N}-PdV+\mu d {\cal N},
\end{equation}
where $\mathcal{E}_N$ is total energy, $P$ is a pressure, $V$ is a volume, $T$ is a temperature, $\cal N$ is a number of electrons, $\mu$ is a chemical potential and $S_N$ is an entropy.
For the total energy per one site $\mathcal{E} = {\mathcal{E}_N}/{N}$ one can write
\begin{equation}
d\mathcal{E} = TdS-Pdv+\mu dn,
\end{equation}
where $v$ is unit cell volume, $n$ is an occupancy. In case $n$ and $v$ are constants the entropy per one site $S(T)=S_N(T)/N$ is
\begin{equation}
dS = \frac{d\mathcal{E}}{T}.
\end{equation}
After integration over temperature in the interval ($T$, $\infty$), one can obtain the entropy as a function of temperature
\begin{equation}
S(T) = S(\infty)-\int\limits_{\mathcal{E}(T)}^{\mathcal{E}(T=\infty)}\frac{d\mathcal{E}(T')}{T'}.
\label{eq:Entropy}
\end{equation}
The entropy at infinite temperature $S(\infty)$ (Entropy of Fermi gas) should be obtained as:  
\begin{equation}
S(\infty) = K\log K -n\log n - (K-n)\log(K-n),
\label{eq:Entropy_inf}
\end{equation}
where $K$ is the site capacity (for single-band model $K = 2$).

General exact expression for total energy of electron system with pair interaction for the single-orbital case in presence of magnetic field $h$  %can be calculated as
is~\cite{Galitskii_Migdal}:
\begin{equation}
\mathcal{E}(T) =\frac{T}{N} \sum_{\mathbf{k} \omega_n \sigma} \left( \varepsilon_{\mathbf{k}} - \sigma h+ \frac{1}{2}\Sigma_{\mathbf{k}\sigma}(\mathrm{i} \omega_{n}) \right) G_{\mathbf{k}\sigma}(\mathrm{i}\omega_{n}) e^{\mathrm{i}\omega_{n}0+},
\label{eq:total_Energy_0}
\end{equation}
where $\varepsilon_{\mathbf{k}}$ is a bare electron spectrum, $\omega_n$ is fermionic Matsubara frequency, $\Sigma_{\mathbf{k}\sigma}(i\omega_n)$ is an electron self-energy, $G_{\mathbf{k}\sigma}(i\omega_n)=(\mathrm{i}\omega_n-\varepsilon_{\mathbf{k}}+\mu-\Sigma_{\mathbf{k}\sigma}(i\omega_n)+\sigma h)^{-1}$ is an exact electron Green function.

It is well known that within infinite dimensions limit the self-energy in DMFT becomes \textbf{k}-independent $\Sigma_{\mathbf{k}\sigma} \to \Sigma_{\sigma}$~\cite{Georges1996}. 
So one can perform summation over $\mathbf{k}$ in \eqref{eq:total_Energy_0} in case $\varepsilon_{\mathbf{k}}$ is excluded from \eqref{eq:total_Energy_0}.
We rewrite the Eq.~\eqref{eq:total_Energy_0}
\begin{equation}
\begin{split}
\mathcal{E}(T)= &T\sum_{\omega_n \sigma}  \left[-1 + (\mathrm{i} \omega_{n} +\mu-\Sigma_\sigma(\mathrm{i}\omega_n) + \sigma h) G_{\sigma}^{\rm loc}(\mathrm{i}\omega_n) \right] + \\
+&\frac{1}{2} T\sum_{n \sigma} \Sigma_\sigma(\mathrm{i}\omega_n) G_{\sigma}^{\rm loc}(\mathrm{i}\omega_n),
\label{eq:total_Energy_Gloc}
\end{split}
\end{equation}
through a local Green function
$G^{\rm loc}_{\sigma}(\mathrm{i}\omega_n) = \frac{1}{N} \sum_{\mathbf{k}} G_{\mathbf{k}\sigma}( \mathrm{i}\omega_n)$.
For fcc lattice the relation $t' = t/2$ results in the occurrence of giant van Hove singularity of density of states (DOS) 
of bare electron spectrum at the bottom of the band.  

In the limit $d = \infty$ there appears a scaling relation of hopping integral, which for fcc lattice is $t = t^\ast/(\sqrt{2}d)$~\cite{Ulmke1998}~($t^\ast$ is constant).
Within DMFT approximation it is natural to take DOS 
\begin{equation}
\rho(E) = \frac1{N}\sum_{\mathbf{k}}\delta(E - \varepsilon_{\mathbf{k}}/t^\ast),
\end{equation}
where $E$ is one-electron level energy in units of $t^\ast$ and
\begin{equation}
\rho(E) = \sqrt{\frac{\theta(1 + \sqrt{2}E)}{\pi (1 +\sqrt{2} E)}}{e^{-(1+\sqrt{2}E)/2}}. 
\label{fccDOS}
\end{equation}
Here $\theta(x)$ stands for Heaviside step function.

We express the local Green function as 
\begin{equation}
G^{\rm loc}_{\sigma}(\mathrm{i}\omega_n) =  \int\limits^{+\infty}_{-\infty} \frac{\rho(E) dE}{\mathrm{i}\omega_n-E +\mu-\Sigma_\sigma(\mathrm{i}\omega_n)+\sigma h}
\label{eq:Gloc}
\end{equation}
through the electron DOS $\rho(E)$. 
Here and below we measure the energy and related quantities in units of $t^\ast$. 

Strong asymmetry of the DOS ensures that the Stoner criterion for the second-order  paramagnet-ferromagnet magnetic phase transition ($ U\rho(E_{\rm f})> 1$) is fulfilled. The existence of itinerant ferromagnetism of the Hubbard model was in details discussed in a number of papers~\cite{Vollhardt1996,Ulmke1998,Vollhardt1999} (also see reference therein).

Finally, analytic expression for bare DOS \eqref{fccDOS} allows one to integrate over the energy in \eqref{eq:Gloc}. Thus we obtain
\begin{multline}
G^{\rm loc}_{\sigma} ( \mathrm{i}\omega_n ) =  \sqrt{\frac{\pi}{\sqrt{2}z_{\sigma}(\mathrm{i}\omega_n)+1}}
\exp\left(-\frac{\sqrt{2}z_{\sigma}(\mathrm{i}\omega_n)+1}{2}\right)\\
\times\left(-\mathrm{i}\cdot\text{sign}\,\text{Im}[z_{\sigma}(\mathrm{i}\omega_n)] + \text{Erfi}\sqrt{\frac{\sqrt{2}z_{\sigma}(\mathrm{i}\omega_n) + 1}{2}}\ \right),
\end{multline}
where Erfi is the imaginary error function and
\begin{equation}
z_{\sigma}(\mathrm{i}\omega_n) = \mathrm{i}\omega_n + \mu -\Sigma_{\sigma}(\mathrm{i}\omega_n) + \sigma h .
\end{equation}

%====================================================================================
\section{Technical details}
\label{sec:technical}
The MFA calculations here were carried out as described in our earlier work~\cite{Igoshev2017}.
For the DMFT calculations we employed the CT-QMC impurity solver~\cite{ctqmc_PRL_2006,ctqmc_PRB_2007,ctqmc_RMP_2011,amulet}.
The DMFT(CT-QMC) computations were done at Monte-Carlo sweeps from $10^5$ to $10^7$ depending on the chosen temperature.
Convergence accuracy of DMFT calculation of occupancy is $10^{-4}$, of total energy is $10^{-5}$ and
of magnetization is $10^{-4}$. Both DMFT and MFA calculations are done at occupancy value $n=0.6$.

To test correctness of our DMFT entropy calculation using equations \eqref{eq:Entropy}, \eqref{eq:Entropy_inf} and \eqref{eq:total_Energy_Gloc} we considered the single-band Hubbard model case on the symmetric bare Bethe lattice and have obtained identical results with ones published before~\cite{Georges1996,Skornyakov2015}.

%====================================================================================
\section{Results and discussion}
\label{sec:results}
\begin{figure}[h]
	\includegraphics[width=1.0\linewidth]{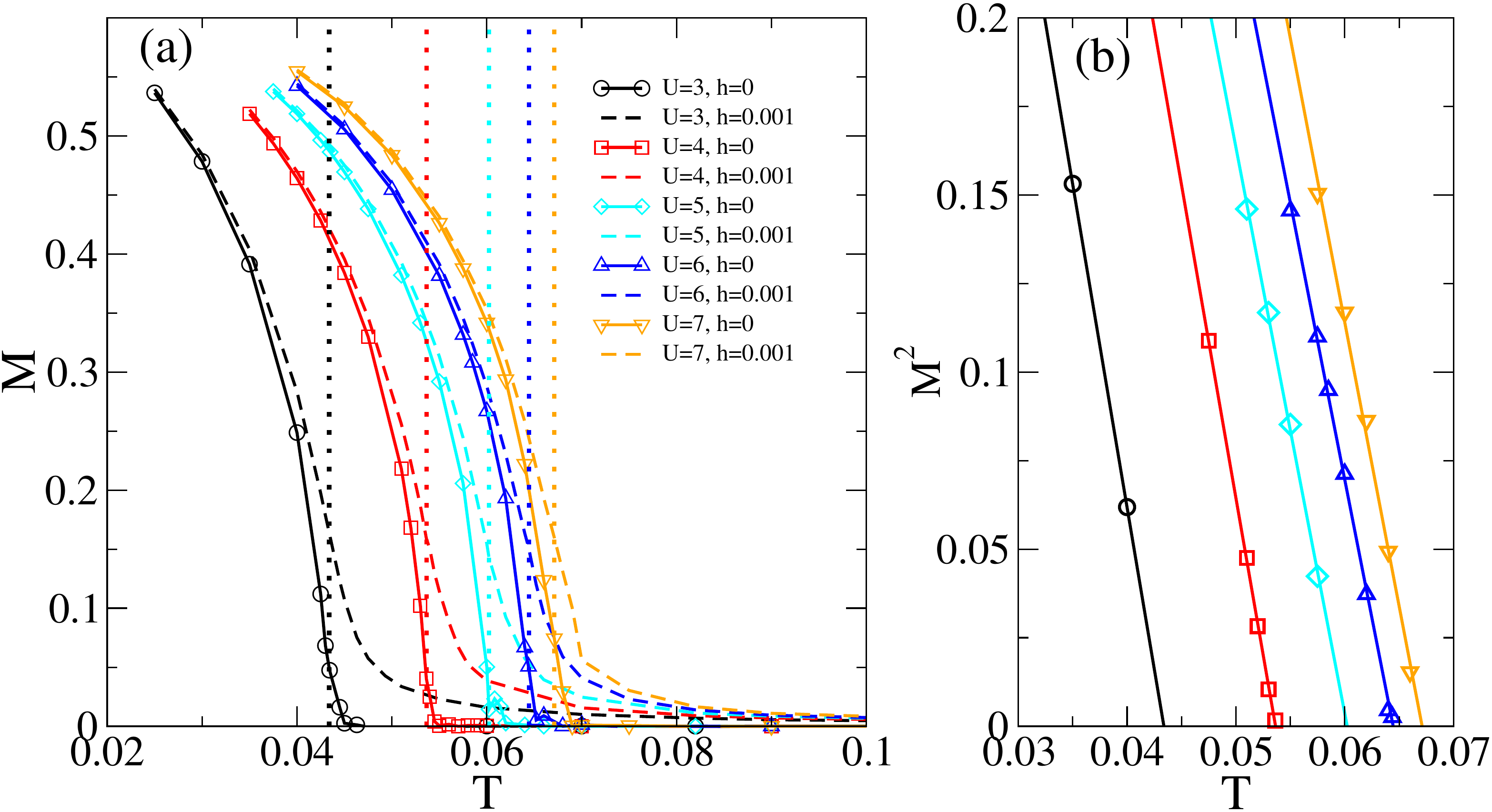}
	\includegraphics[width=1.0\linewidth]{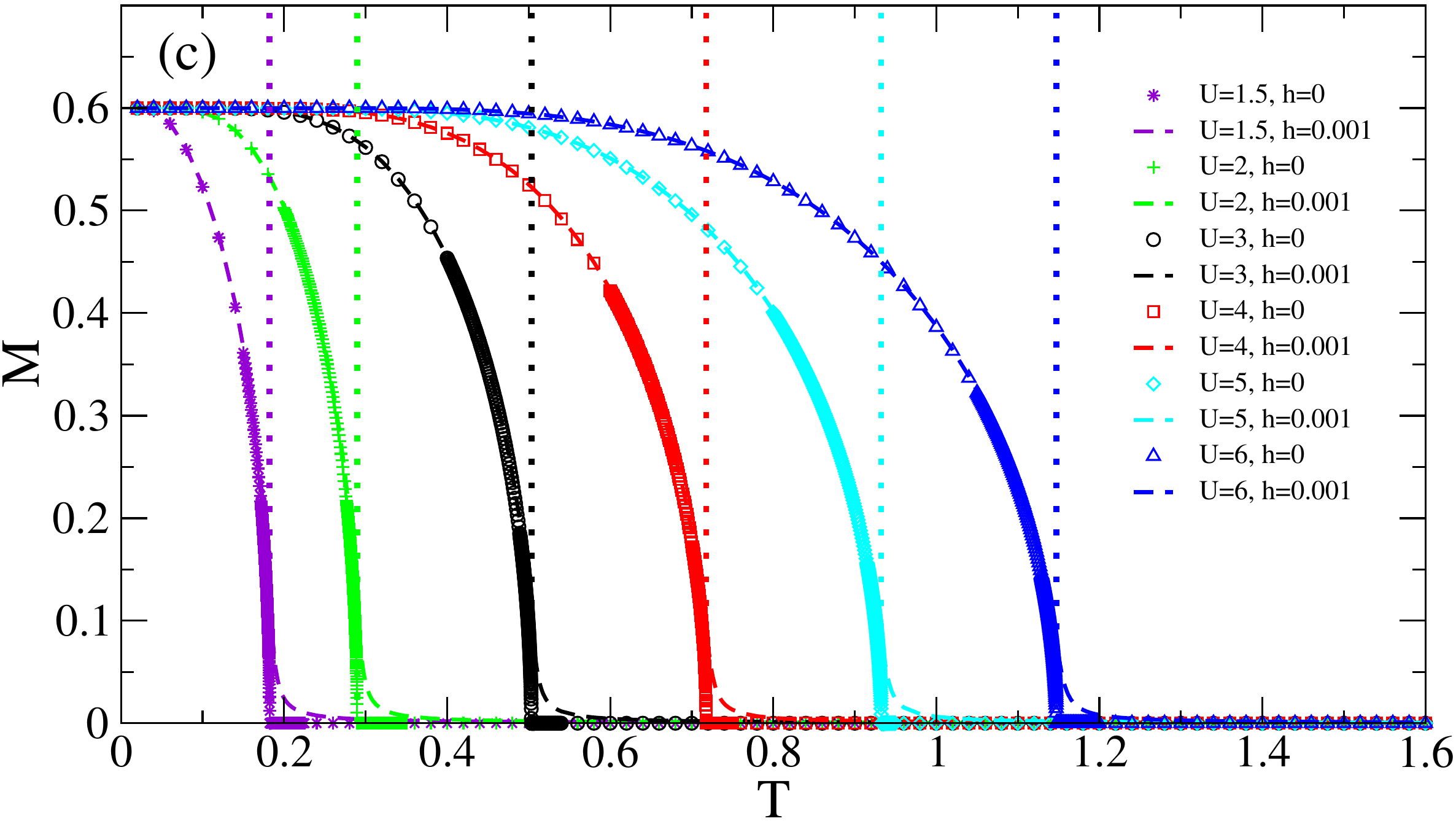}
	\caption{DMFT (panel (a)) and MFA (panel (c)) calculated magnetization $M$ vs temperature $T$ for single-band Hubbard model for infinite-dimensional fcc lattice with different $U$ values (solid lines corresponds to $h=0$, dashed lines to  $h=0.001$). Vertical dotted lines correspond to $T_C$; Panel (b): $M^2$ for the case of DMFT: dots corresponds to DMFT results, lines to least square method fit. Occupancy $n=0.6$.}
	\label{fig:mag}
\end{figure}

\begin{figure}[h]
	\includegraphics[width=1.0\linewidth]{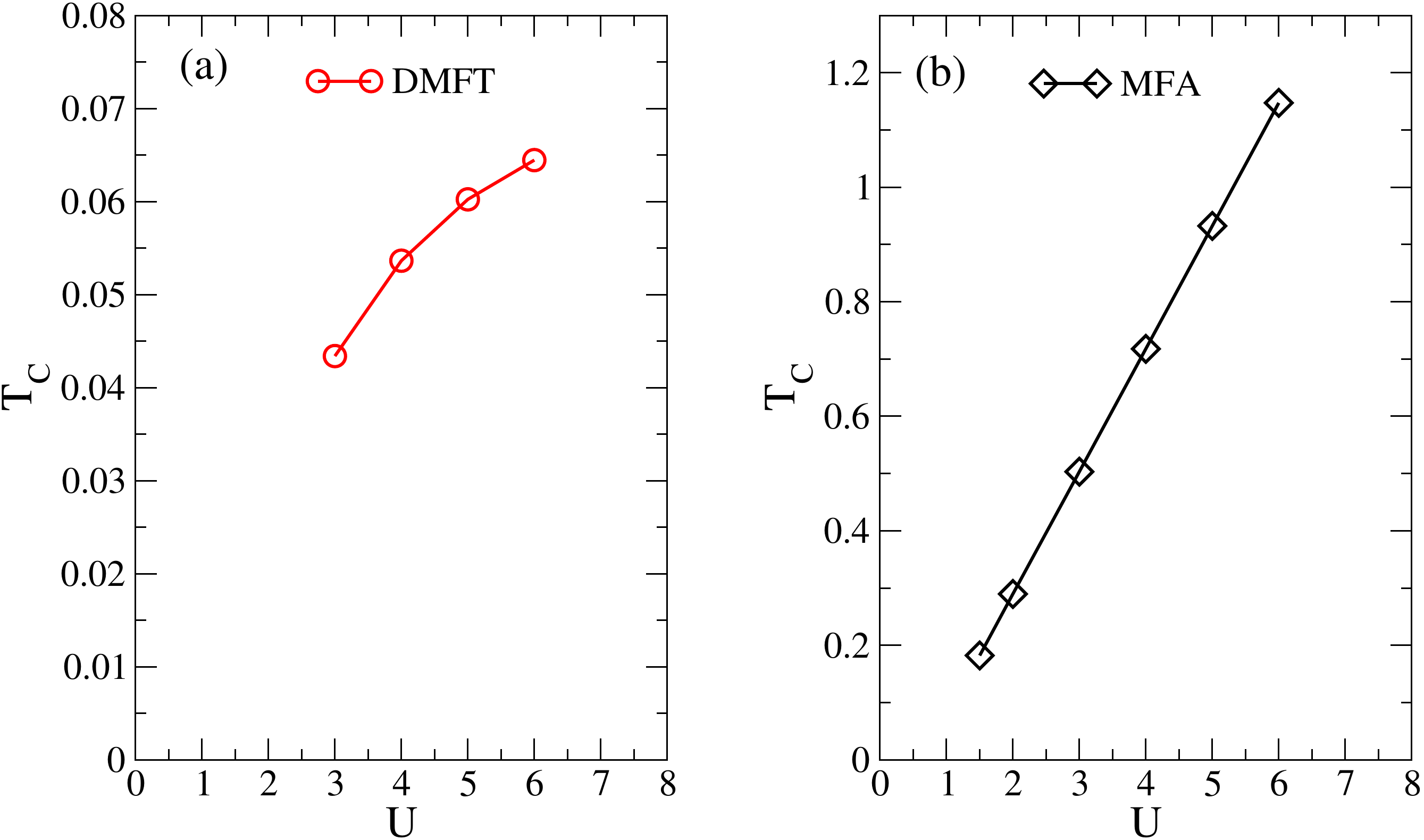}
	\caption{Curie temperature $T_C$ dependence on value of the Coulomb interaction $U$ obtained within DMFT (panel (a)) and MFA (panel (b)).}
	\label{fig:Tc_U}
\end{figure}

Since MCE has its strongest manifestation at magnetic phase transition, first of all one should precisely define the value of Curie temperature $T_C$
for both MFA and DMFT.
Since it is numerically difficult to get down to zero $M$ in the ferromagnetic phase with increase of temperature with required accuracy, the Curie temperature is determined using analytical magnetization behavior near the critical temperature $M \sim \sqrt{T-T_{C}}$ for the second order phase transition. % in the ferromagnetic phase. This is done as 
Solid lines of figure~\ref{fig:mag}(a) show the magnetization obtained within DMFT for single-band Hubbard model with infinite-dimensional fcc lattice 
for different $U$ values. The $T_C$ values correspond to the point where $M^2$ linearly goes to zero. To find it $M^2$ is fitted with least square method to a linear function as plotted in Figure~\ref{fig:mag}(b).
One can see that with an equal step increase in value of the Coulomb interaction, the $T_C$ values for different $U$ become closer to each other for DMFT.
Corresponding curve $T_C$ versus $U$ is shown in Fig. \ref{fig:Tc_U}(a).

Let's compare DMFT results for magnetic moment and $T_C$ with an MFA solution (see Fig.~\ref{fig:mag}(c)). 
For the MFA case $T_C$ can be obtained directly from $M$ curve since numerical accuracy that can be achieved for the MFA calculations is orders of magnitude better than for DMFT. 
Immediately one can notice that MFA $T_C$ values are order of magnitude larger than those for DMFT. 
It corresponds to the well known fact that MFA strongly overestimates $T_C$.
In the case of MFA $T_C$ linearly depends on value of the Coulomb interaction $U$ (see Fig. \ref{fig:Tc_U}(b)) which is not typical behaviour ~(square root at moderate $U$ and constant at very large $U$~\cite{Moriya}) and caused by the vicinity of Fermi level to giant van Hove singularity of DOS at the bottom of the band breaking the applicability of low temperature Sommerfeld expansion. 
Within the DMFT the quantum fluctuation weakens the influence of van Hove singularity and the Curie temperature dependence is square root like. 

%{\color{red} It happens since DOS has square root singularity at the bottom of the band and thus here low temperature Sommerfeld expansion does not work.}

\begin{figure}[h]
	\includegraphics[width=1.0\linewidth]{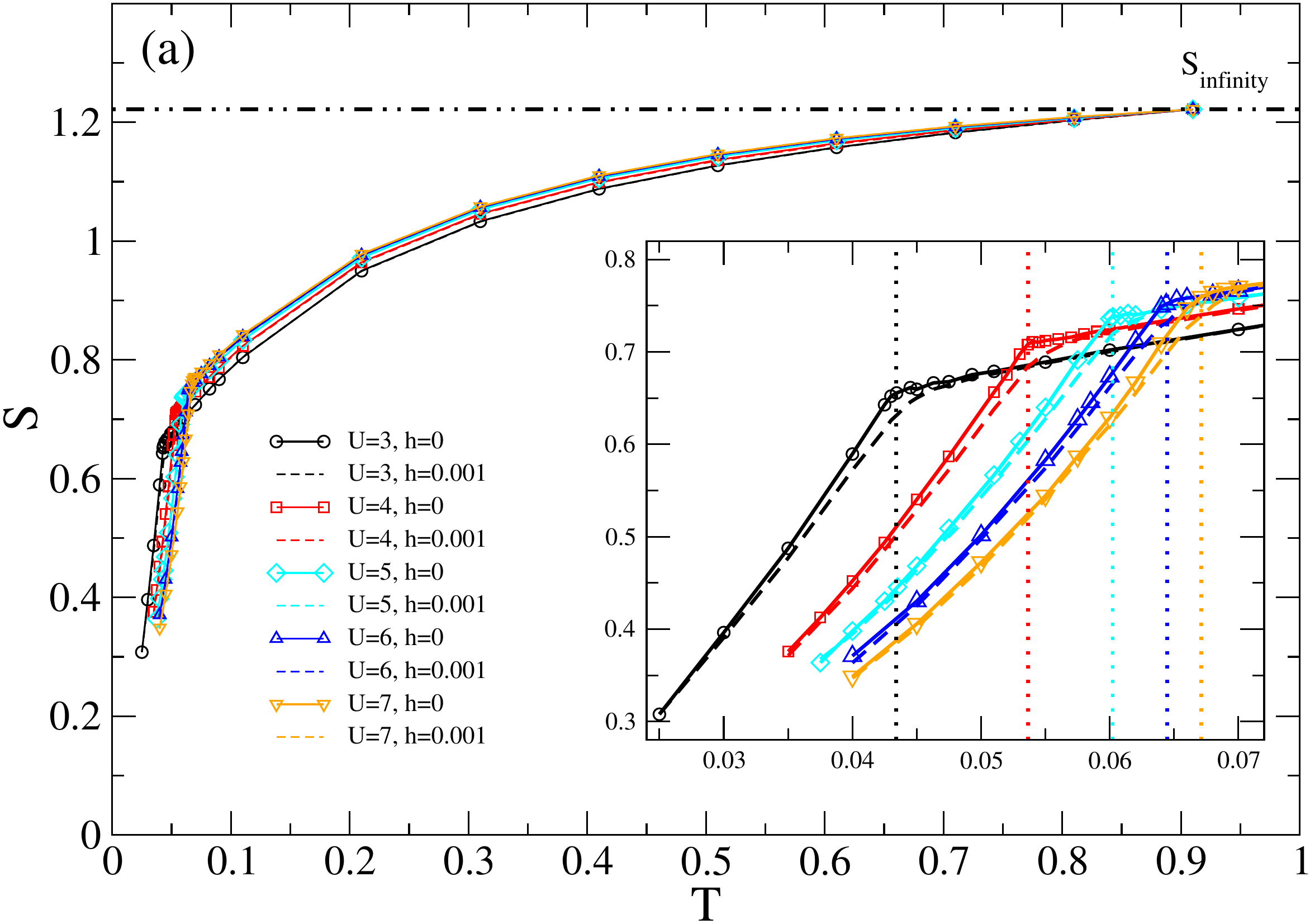}
	\includegraphics[width=1.0\linewidth]{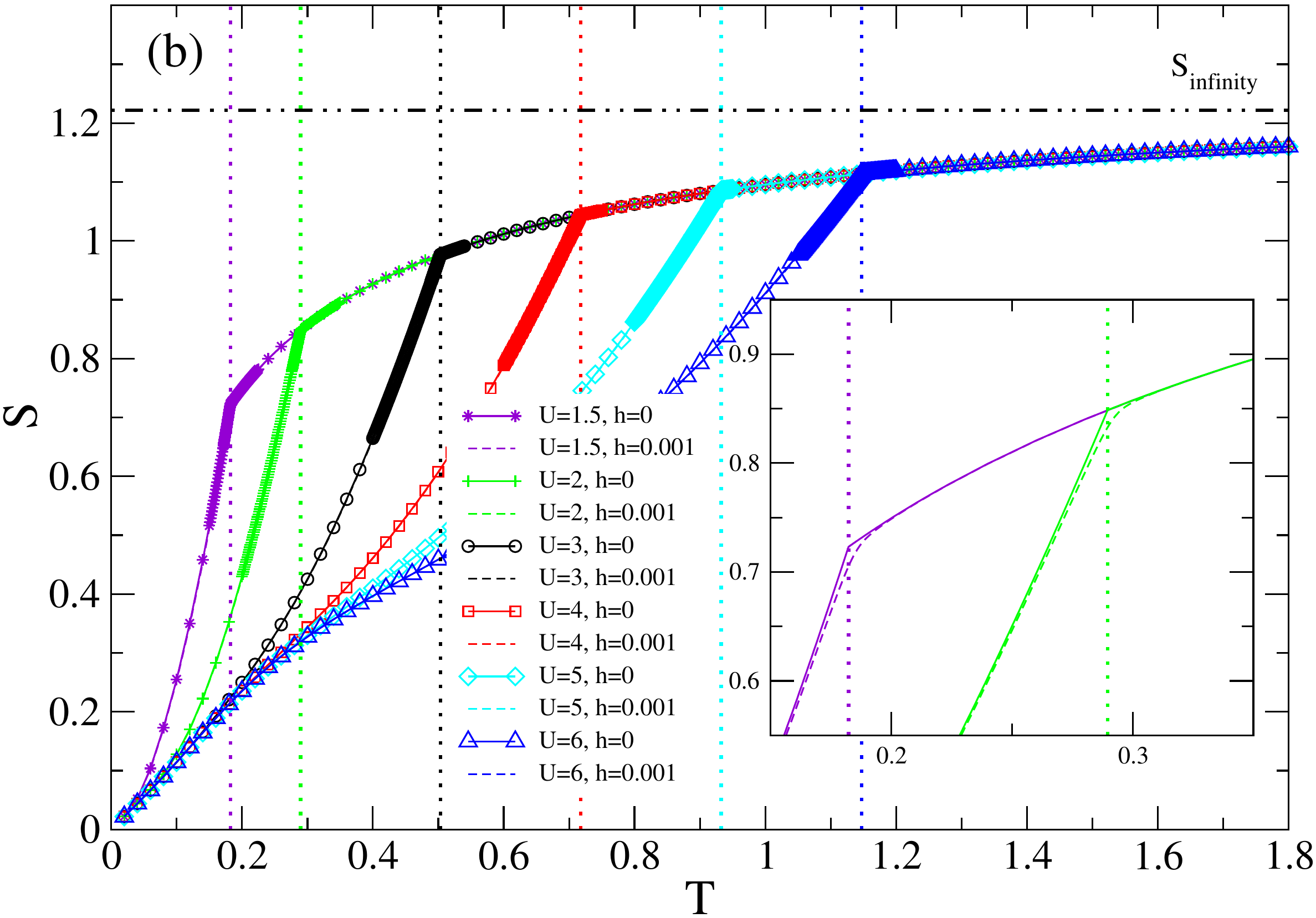}
	\caption{Entropy $S$ vs temperature $T$ within DMFT (a) and MFA (b) for single-band Hubbard model (infinite-dimensional fcc lattice) at different $U$ values. 
	Solid lines corresponds to $h=0$, dashed lines to $h=0.001$.
	Occupancy is $n=0.6$. Vertical dotted lines correspond to $T_C$.
	Insets show vicinity of the entropy kinks.}
	\label{fig:S}
\end{figure}

The DMFT (solid lines at panel (a) of Fig. \ref{fig:S}) and MFA (solid lined at panel (b) of Fig. \ref{fig:S}) entropy $S$ dependence on temperature $T$ at different values of the Coulomb interaction $U$ for zero magnetic field is shown.
Both in the case of the DMFT and in the case of the MFA results the entropy has a clearly visible kink at the $T_C$ (see insets of Fig. \ref{fig:S}(a) and Fig. \ref{fig:S}(b)). The value of entropy corresponding to the kink grows up with increase of $U$ value for DMFT and MFA. However the entropy above $T_C$ for the DMFT is different for different $U$ values while for the MFA solution paramagnetic $S$ goes along the same line.
At $T=0.91$ the value of DMFT calculated entropy is already very close to the $S(\infty)$ obtained from Eq.~\eqref{eq:Entropy} while MFA for that requires much higher temperatures.
For the finite magnetic field ($h=0.001$, dashed lines on Fig. \ref{fig:S}(a,b)) value of $S$ is less than for $h=0$ and kink in entropy is smeared out.

\begin{figure}[h]
	\includegraphics[width=1.0\linewidth]{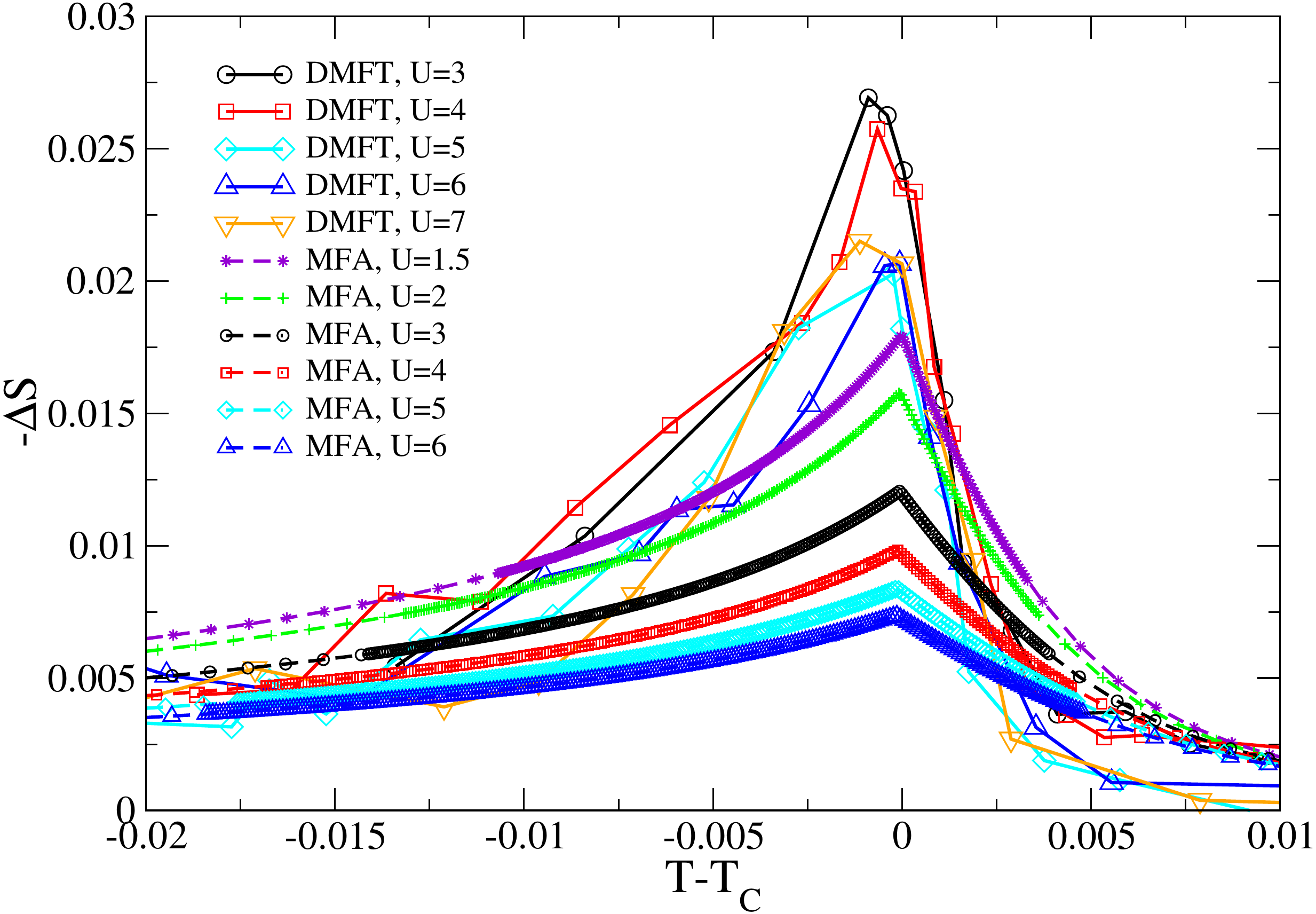}
	\caption{MCE isothermal entropy change $\Delta S (T)$ within the magnetic field change $\Delta h=0.001$ calculated within DMFT (solid lines) and MFA (dashed lines) for single-band Hubbard model (infinite-dimensional fcc lattice). Occupancy is $n=0.6$. Zero corresponds to $T_C$.}
	\label{fig:dS}
\end{figure}

In Figure~\ref{fig:dS} one of the MCE potentials, isothermal entropy change $\Delta S(T)$, under magnetic field change $\Delta h = h =0.001$ obtained from DMFT (solid lines) and MFA (dashed lines) results of Fig.~\ref{fig:S} is shown. For both methods maximum value of $\Delta S (T)$  ($\Delta S_{\rm max}$) is observed at the Curie temperature $T_C$. In the MFA $\Delta S_{\rm max}$ quite significantly decreases as $U$ grows~\cite{Igoshev2017}. Similar but much weaker decrease of $\Delta S_{\rm max}$ is found for DMFT results. For DMFT case at $U$=5, 6 and 7 $\Delta S_{\rm max}$ practically does not depend on $T$.
Let us note that for DMFT $\Delta S_{\rm max}$ is slightly larger than for MFA. A peak width of $\Delta S (T)$ at half height is approximately the same for both methods. 

We have also estimated the relative cooling power (RCP) given by:
\begin{equation}
R C P=-\Delta S_{\rm max} \times \delta T_{F W H M}
\end{equation}
where
$\delta T_{F W H M}$
is the full-width at half maximum of $\Delta S_{\rm max}$ versus
temperature, to evaluate the magnetic cooling efficiency. The RCP values correspond to the amount of heat
transferred between the cold and hot sides in an ideal refrigeration cycle. Despite $\Delta S_{\rm max}$ is bigger in DMFT than in MFA the RCP values for the same $U$ are very close to each other and decay while $U$ grows.

%====================================================================================
\section{Conclusion}
\label{sec:conclution}

In this paper we reported a comparative study of MCE within mean-field (Stoner) approximation (MFA) and within dynamical mean-field theory (DMFT) for partially filled single-band Hubbard model on infinite-dimensional face-centered cubic lattice.
To this end MFA and DMFT calculations
of temperature dependence of magnetization, total energy and finally entropy for a set of Coulomb interactions $U$ at zero and finite values of magnetic field $h$ were performed. As expected DMFT corrects Curie temperature values compared to MFA.

We have found that despite entropy as a function of temperature has some differences between DMFT and MFA the MCE entropy change $\Delta S(T)$ has quite similar behavior for both methods. We believe that it could be explained by the fact that DMFT is a mean-field theory for magnetization for the case of second order magnetic phase transition~\cite{Byczuk2002}. 
A qualitative difference of the results of these approximations is suppression of the influence of van Hove singularity of DOS onto thermodynamic properties by local quantum fluctuations, which restores square-root-like dependence of $T_C$ on $U$.    
This also explains the dramatic difference of value of $T_C$ within these approximations. 

Direct comparison of DMFT and MFA $\Delta S(T)$ shows that $\Delta S_{\rm max}$ is 2-4 times higher in DMFT than in MFA while width of $\Delta S(T)$ is approximately the same. However relative cooling power (RCP) for the same model parameters is very similar for DMFT solution and MFA one and decreases while $U$ grows.

%====================================================================================
\section{Acknowledgements}
This work was supported by UB RAS programm No. 18-2-2-1.
NSP thanks President of Russia grant for young scientists No. MK-1683.2019.2.
PNA thanks the state assignment of Minobrnauki of Russia (theme ``Electron'' No. AAAA-A18-118020190095-4).
Part of the CT-QMC computations were performed at
``URAN'' supercomputer of the Institute of Mathematics and Mechanics of the RAS Ural Branch.

\end{document}